\documentstyle[preprint,epsfig,aps]{revtex}
\tightenlines
\begin{document}
\draft

\preprint{MKPH-T-99-4}
\title{Medium Effects in Coherent Pion Photo- and Electroproduction on
 $^4He$ and $^{12}C$ }

\author{ D. Drechsel and L. Tiator}
\address{Institut f\"ur Kernphysik, Universit\"at Mainz, 55099 Mainz, Germany}

\author{ S. S. Kamalov\footnote
{Permanent address: Laboratory of Theoretical Physics, JINR Dubna,
Head Post Office Box 79, SU-10100 Moscow, Russia}
and Shin Nan Yang}
\address{Department of Physics, National Taiwan University, Taipei, Taiwan
10634, Republic of China}

\date{\today}
\maketitle

\begin{abstract}
  Coherent $\pi^0$ photo- and electroproduction on $^4$He and $^{12}$C
  nuclei is investigated in the framework of a distorted wave impulse
  approximation in momentum space. The elementary process is described
  by the recently developed unitary isobar model.  Medium effects are
  considered by introducing a phenomenological $\Delta$ self-energy.
  The recent experimental data for $^4$He and $^{12}$C can be well
  described over a wide range of energies and emission angles by the
  assumption that the $\Delta$-nuclear interaction saturates.
\end{abstract}


\section{INTRODUCTION}

Recently, much attention has been paid to experimental investigations
of coherent $\pi^0$ photo- and electroproduction off nuclei, i.e.
\begin{equation}
 \gamma^* + A(gs) \rightarrow  A(gs) + \pi^0\,,
\end{equation}
where $\gamma^*$ is a real or virtual photon and $A(gs)$ is a nucleus
in its ground state.  This reaction is of special interest for nuclei
with zero spin and isospin. In this case the theoretical treatment
simplifies and contains only a minimum number of ingredients.  Since
the description of the nuclear ground state is well under control, the
reaction is then suitable for analyzing medium effects in the production
and propagation of $\Delta$ resonance.

The earliest theoretical and experimental studies of this reaction had
the aim to extract additional information about the elementary
amplitude, in particular about the values of the $M_{1+}$ and $M_{1-}$
multipoles, which are related to the excitation of the resonances
$\Delta(1232)$ and $N^{\ast} (1440)$, respectively. The general
interest in this field was revived in the 80's in the context of
studying the photoproduction of neutral pions on the proton in order
to check the threshold predictions of low energy theorems.

On the other hand, coherent $\pi^0$ photoproduction off nuclei was
considered as unique tool to examine the properties of the
pion-nucleus interaction and to obtain information about the behavior
of the pion in the nuclear medium. The theoretical method used for
this purpose was the distorted wave impulse approximation (DWIA).
While this method was originally formulated in coordinate space with a
purely phenomenological optical potential, the modern approach to DWIA
uses a representation in momentum space.  The first application of the
DWIA in the momentum space to pion photoproduction on nuclei was due
to Eramzhyan and collaborators \cite{RAE1,RAE2}.  Within this approach
the pion-nucleus interaction (or final state interaction -- FSI) is
treated dynamically starting from the elementary pion-nucleon
amplitude, and consistently taken into account of nonlocalities of the
photoproduction operator and off-shell effects.  In particular, this
method was applied to the case of coherent $\pi^0$ production by Ref.
\cite{RAE3}.  In the same spirit, the authors of Refs.
\cite{Yang,Tanabe,NBL} developed their dynamical model for pion
photoproduction on the nucleon.

However, FSI is only one aspect of the medium effects due to the
modification of the pion propagator in the nuclear medium.  Another
important issue is the modification of the resonance characteristics
(width and position) inside a nucleus. This problem was the subject of
numerous investigations, especially in the framework of the
$\Delta$-hole model.  For a review and further literature, the reader
is referred to the monographs~\cite{EK,Weise,Koch2}.  This issue will
be the main subject of our study.  For this purpose we apply, for the
first time, the Unitary Isobar Model (UIM) of Ref.~\cite{UIM} to pion
photoproduction on nuclei. This recently developed model is
particularly adopted to describe the elementary process on the nucleon
in the nuclear environment.  An important advantage of this model is
that the contributions of the background (Born terms and vector meson
exchange) and "dressed" $\Delta$ resonance are separated in a unitary
way. As we will show below this allows us to incorporate medium
effects in a self-consistent way with regard to both the "bare"
$\gamma N \Delta$ vertices and $\Delta$ excitations due to pion
rescattering. Another advantage of the UIM is that it covers both pion
photoproduction and pion electroproduction up to 4-momentum transfers
of $Q^2=-k^2 \simeq 4\,(GeV/c)^2$. Therefore, this model provides
quite a unique opportunity to investigate the $Q^2$ dependence of
medium effects. Note that the first theoretical study of the medium
effects in the coherent $\pi^0$ electroproduction on the heavy nuclei
was done in Ref.\cite{Hirenzaki}.

The structure of this paper is as follows. In sections 2 and 3 we
consider the main theoretical ingredients to describe pion photo- and
electroproduction on nuclei.  Our results and predictions are
presented in section 4.  We shall demonstrate that all the available
data for coherent pion photoproduction on $^4He$ and $^{12}C$ can be
described, in a self-consistent way, with the same parameter set for
the $\Delta$ self-energy. Finally, our conclusions are summarized in
section 5.

\section{General expressions}

\subsection{Differential cross sections}

In the case of an unpolarized beam and unpolarized target,
the 5-fold differential
cross section in the $lab$ frame can be written as
(see Ref.\cite{DT} for details)
\begin{equation}
\frac{d\sigma}{d\Omega_{e'}\,dE_{e'}\,d\Omega_{\pi}} =
\Gamma \frac{d\sigma_V}{d\Omega_{\pi}}\,,
\end{equation}
which defines the virtual photon cross section

\begin{equation}
\frac{d\sigma_V}{d\Omega_{\pi}} = \frac{d\sigma_T}{d\Omega_{\pi}}
+ \epsilon \frac{d\sigma_L}{d\Omega_{\pi}}
+ \sqrt{2\epsilon (1+\epsilon)}\frac{d\sigma_{TL}}{d\Omega_{\pi}}\,
  \cos{\Phi_{\pi}}
+ \epsilon\frac{d\sigma_{TT}}{d\Omega_{\pi}}\,\cos{2\Phi_{\pi}}\,,
\end{equation}
with $\epsilon$ and $\Gamma$ the degree of transverse
polarization and
the flux of the virtual photon, respectively. Denoting the photon's
$lab$ energy by $\omega_L=E_{e}-E_{e'}$ and its
three-momentum by ${\bf k}_L$, we have

\begin{equation}
\epsilon=\left[ 1 + 2\frac{{\bf k}^2_L}{Q^2}\tan^2 \frac{\theta_e}{2}
\right]^{-1}\,,\qquad
\Gamma = \frac{\alpha}{2\pi^2}\ \frac{E_{e'}}{E_{e}}\,\frac{K}{Q^2}\
\frac{1}{1-\epsilon}\,.
\end{equation}

In accordance with our previous work, we define the flux by the photon
"equivalent energy" $K=(s_A-M^2_A)/2M_A$, with the
Mandelstam variable $s_A=E^2=-Q^2+M^2_A+2\omega_{L} M_A$, $M_A$ the
mass of the nuclear target, and the square of the four-momentum
transfer $Q^2=-k^2={\bf k}^2_L - \omega^2_L >0$.

The first two terms in Eq. (3) are the transverse (T) and longitudinal
(L) cross sections, and the third and fourth terms are the
transverse-longitudinal (TL) and transverse-transverse (TT)
interference cross sections.  The latter ones can be separated by use
of their typical dependence on the azimuthal angle $\phi_{\pi}$ of
pion emission.  As in the case of pion electroproduction on the free
nucleon, these four cross sections may be expressed in terms of the
nuclear tensor $W_{\mu\nu}$, with $\mu$ and $\nu$ corresponding to the
Cartesian coordinates $x, y$ or $z$, which leads to the following
relations in the $cm$ frame:

\begin{mathletters}
\begin{equation}
\frac{d\sigma_T}{d\Omega_{\pi}}=\frac{k_{\pi}}{2k_{\gamma}^{cm}}
(W_{xx}+W_{yy}),\quad\quad
\frac{d\sigma_{TT}}{d\Omega_{\pi}}=\frac{k_{\pi}}{2k_{\gamma}^{cm}}
(W_{xx}-W_{yy}),
\end{equation}
\begin{equation}
\frac{d\sigma_L}{d\Omega_{\pi}}=\frac{k_{\pi}}{k_{\gamma}^{cm}}
 \frac{Q^2}{\omega_{\gamma}^2}\,W_{zz},
\quad\quad
\frac{d\sigma_{TL}}{d\Omega_{\pi}}=-\frac{k_{\pi}}{k_{\gamma}^{cm}}
\frac{Q}{\omega_{\gamma}}\,Re\,W_{xz}\,,
\end{equation}
\end{mathletters}
with $k_{\gamma}^{cm}=(E^2-M_A^2)/2E$ the "photon equivalent energy"
in the $cm$ frame and $k_{\pi}=\mid
{\bf k}_{\pi} \mid$ the pion momentum. Note that we use a right-handed
coordinate system with the positive $z$-axis along the virtual photon
momentum ${\bf k}_{\gamma}$ and the $y$-axis along the vector $[{\bf
  k}_{\gamma}\times{\bf k}_{\pi}\,]$.
The nuclear tensor $W_{\mu\nu}$ is defined in terms of the
nuclear transition amplitudes $F_{fi}^{(\mu)}$ describing the
transitions from the initial states $\mid i >=\mid J_i M_i>$
to the final states $\mid f >=\mid J_f M_f>$ with spin $J_{i(f)}$ and
projection $M_{i(f)}$. After summing and averaging over the nuclear
spin degrees of freedom we obtain
\begin{equation}
W_{\mu\nu}=\frac{1}{2J_i+1}Re\sum_{M_i,M_f}F_{fi}^{(\mu)}F_{fi}^{(\nu)*}\,.
\end{equation}

For the purpose of the numerical calculations it is convenient to express
the nuclear
amplitudes in the covariant spherical basis
${\bf e}=\{{\bf e}_{+1},{\bf e}_{-1},{\bf e}_{0}\}$ by the relations
\begin{equation}
F_{fi}^{(\lambda)}=-\frac{\lambda}{\sqrt{2}}(F_{fi}^{(x)}+
i\lambda F_{fi}^{(y)}),\quad\quad F_{fi}^{(0)}=F_{fi}^{(z)}\,,
\end{equation}
where  $\lambda=\pm 1$  is the helicity of the transverse photon.

In the case of targets with zero spin, the general symmetry properties
and the pseudoscalar nature of the pion have the consequence that the
transverse nuclear amplitude $F_{fi}^{(\lambda)}$ contains only a term
proportional to $[{\bf k}_{\pi}\times {\bf k}_{\gamma}]$ in the
pion-nucleus $cm$ frame. Hence
\begin{equation}
F_{fi}^{(\lambda)}({\bf k}_{\pi},{\bf k}_{\gamma};Q^2)=
F_0({\bf k}_{\pi},{\bf k}_{\gamma};Q^2)\,
[\hat{\bf k}_{\pi}\times \hat{\bf k}_{\gamma}] \cdot {\bf e}_{\lambda}\,,
\qquad
F_{fi}^{(0)}=0\,,
\end{equation}
where $\hat{\bf k}_{\pi}$ and $\hat{\bf k}_{\gamma}$ are unit vectors
in the directions of the pion and the virtual photon momentum,
respectively.

In the right-handed coordinate system with the $z$-axis along
$\hat{\bf k}_{\gamma}$ and the $y$-axis along $[\hat{\bf
  k}_{\gamma}\times \hat{\bf k}_{\pi}\,]$, only the $W_{yy}$ tensor
component remains. Therefore,
$d\sigma_T/d\Omega_{\pi}=-d\sigma_{TT}/d\Omega_{\pi}$ and all other
cross sections in Eq. (3) vanish. In this case the expression for the
virtual photon cross section simplifies to~\cite{Hirenzaki}
\begin{equation}
\frac{d\sigma_V}{d\Omega_{\pi}} = \frac{d\sigma_T}{d\Omega_{\pi}}\,
(1-\epsilon\,\cos{2\Phi_{\pi}})\,.
\end{equation}

It should be noted that Eqs. (8-9) are general.
As a consequence the differential cross section has a
$\sin^2{\theta_{\pi}}$ and $(1-\epsilon\,\cos{2\Phi_{\pi}})$
dependence in all cases, independently of the reaction mechanism.
In fact, this peculiarity of coherent pion photo- and electroproduction
on spin zero nuclei is often used to separate coherent and incoherent
contributions, or to measure the degree of photon polarization,
because the photon asymmetry $\Sigma=-d\sigma_{TT}/d\sigma_T$
equals unity.

\subsection{PWIA and DWIA nuclear amplitudes}

Let us first consider the nuclear amplitude in a simple
plane wave impulse approximation (PWIA). The corresponding
amplitude is denoted by $V_{fi}^{(\lambda)}$.
It can be expressed in terms of the elementary pion
electroproduction amplitudes $f_{\gamma \pi}^{(\lambda)}$
and reduced to the form
\begin{equation}
V_{fi}^{(\lambda)}({\bf k}_{\pi},{\bf k}_{\gamma};Q^2)= {\cal W}_A
\int d{\bf r}\,\Psi_{f}^{*}({\bf r}) \sum_{j=1}^A
e^{i{\bf q}\cdot{\bf r}_j}\,f_{\gamma \pi}^{(\lambda)}({\bf k}_{\pi},
{\bf k}_{\gamma},{\bf p}_j;Q^2)\,\Psi_{i}({\bf r})
\end{equation}
with ${\bf r}=\{{\bf r}_1,{\bf r}_2,...,{\bf r}_A\}$ the set
of nucleon coordinates, a phase space factor ${\cal W}_A$ given by
\begin{equation}
{\cal W}_A =\frac{W}{E}\sqrt{\frac{E_A(k_{\pi})E_A(k_{\gamma})}
{E_N(p\,')E_N(p)}}\,,
\end{equation}
${\bf p}$ and ${\bf p}'$ the initial and final nucleon momenta,
$E_{A(N)}$ the nuclear (nucleon) energies, ${\bf q}={\bf k}_{\gamma}-
{\bf k}_{\pi}$ the momentum transfer to the nucleus, and
$Q^2 = {\bf k}^2_{\gamma} - \omega^2_{\gamma}$ expressed by the $cm$
momentum and energy of the virtual photon.

The invariant amplitude $f_{\gamma \pi}^{(\lambda)}$ describes the
elementary process on the free nucleon. It is evaluated within the
framework of the UIM in terms of the standard CGLN amplitudes
$F_i(W,\cos{{\tilde \theta}_{\pi}},Q^2)$, with the pion angle
${\tilde \theta}_{\pi}$ and the total energy  $W$
in the $\gamma N$ $cm$ system,
\begin{equation}
W=\sqrt{(\omega_{\gamma}+E_N({\bf p}))^2-({\bf k}_{\gamma}+{\bf p})^2}\,.
\end{equation}
Note that in the literature there are several prescriptions for the
relations between total pion-nuclear energy $E$ and $W$ with various
physical or mathematical motivations (see for example
Refs.\cite{Landau,Mach,Gurv,Chumb,KTB}).  Using this freedom we in
principle can effectively take into account medium effects and improve
by this way (or "optimize") impulse approximation. However, our main
aim is the study the medium effects explicitly. Therefore, in the
present work we will use physically transparent relation (12) which
does not contain information about residual nucleus and about final
state.

The next problem is related to the treatment of Fermi motion, i.e. the
dependence of the amplitudes on the nucleon's momentum ${\bf p}$.
Previous studies of pion-nucleus scattering \cite{Landau,Mach} and
pion photoproduction reactions \cite{TRD,EGK} showed that a good
approximation to the proper averaging over Fermi motion is given by
the "factorization approximation", which supposes that the main part
of Fermi motion can be accounted for by evaluating the elementary
amplitude at "effective" nucleon momenta ${\bf p}$ and ${\bf p'}$ in
the initial and final states,
\begin{equation}
  {\bf p} =-\frac{{\bf k}_{\gamma}}{A} - \frac{A-1}{2A}{\bf
    q}\,,\quad\quad {\bf p}\,'=-\frac{{\bf k}_{\pi}}{A} +
  \frac{A-1}{2A}{\bf q}\,,
\end{equation}
where $A$ is the number of nucleons. The same approximation, but in
the nuclear rest frame is taken in Refs.  \cite{Boffi,Carrasco}.  Note
that this approximation is based on the fact that for Gaussian nuclear
wave functions (which reproduce the ground state of light nuclei
sufficiently well) the replacement leads to an exact treatment of the
terms linear in ${\bf p}/2M$ in the elementary amplitude. Moreover, it
allows a consistent description of nuclear and nucleon kinematics, by
means of a simultaneous conservation of energies and momenta for both
the pion-nucleus and the pion-nucleon systems. As an example this
factorization approximation was tested numerically for pion
photoproduction on p-shell nuclei \cite{EGK}. Recently this
approximation was also examined within a more elaborate relativistic
model for coherent pion photoproduction on $^{12}C$ \cite{Mosel}.

In the case of coherent photo- and electroproduction of neutral pions on
spin and isospin zero nuclei, the factorization approximation allows us to
obtain the following simple expression for the PWIA amplitude
$V_{fi}^{(\lambda)}$
\begin{equation}
V_{fi}^{(\lambda)}({\bf k}_{\pi},{\bf k}_{\gamma};Q^2)=\,A\,{\cal W}_A \,
f_2({\bf k}_{\pi},{\bf k}_{\gamma};Q^2)\,F_A(q)\,
[\hat{\bf k}_{\pi}\times \hat{\bf k}_{\gamma}] \cdot {\bf e}_{\lambda}\,,
\end{equation}
where $f_2=[f_2(p\pi^0)+f_2(n\pi^0)]/2$ is the isoscalar non spin-flip
part of the elementary amplitude boosted to the pion-nucleus $cm$
frame.  The nuclear form factor is normalized to $F_A(0) = 1$. It can
be extracted from the nuclear charge form factor by the relation
$F_{A}^{ch}(q)=F_{A}(q)F_p^{ch}(q)$, where $F_p^{ch}(q)$ is the proton
charge form factor. The nuclear form factor $F_A(q)$ used in our
present work is taken from Ref.\cite{RAE3}.

Let us now consider the full DWIA amplitude including pion
distortion or FSI effects. In momentum space it takes the form
~\cite{RAE1}
\begin{equation}
 F_{fi}^{(\lambda)}({\bf k}_{\pi},{\bf k}_{\gamma};Q^2) =
 V_{fi}^{(\lambda)}({\bf k}_{\pi},{\bf k}_{\gamma};Q^2)+
 D_{fi}^{(\lambda)}({\bf k}_{\pi},{\bf k}_{\gamma};Q^2)\,,
\end{equation}
where $ D_{fi}^{(\lambda)}$ is the contribution from the pion-nucleus
interaction expressed in terms of the pion-nucleus elastic scattering
amplitude $F_{\pi A}$,
\begin{equation}
 D_{fi}^{(\lambda)}({\bf k}_{\pi},{\bf k}_{\gamma};Q^2)=
-\frac{a}{(2\pi)^2} \sum_{M'_f} \int \frac{ d {\bf k}_{\pi}'}
{{\cal M}(k_{\pi}')}\frac{ F_{\pi A}({\bf k}_{\pi},{\bf k}_{\pi}')\,
 V_{fi}^{(\lambda)}({\bf k}_{\pi}',{\bf k}_{\gamma};Q^2)}
{E(k_{\pi}) - E(k_{\pi}') + i\epsilon}
\, \, .
\end{equation}
In this equation, the relativistic reduced mass of the pion-nucleus
system is given by ${\cal M}(k_{\pi})=\omega_{\pi}(k_{\pi})E_{A}
(k_{\pi})/E(k_{\pi})$. Note that the main difference compared to
standard DWIA is the factor $a=(A-1)/A$, which eliminates double
counting of pion rescattering on one and the same nucleon. Such
effects are in fact already included in the elementary amplitude.
Finally, the pion scattering amplitude $F_{\pi A}$ is constructed as
solution of a Lippmann-Schwinger integral equation (for details see
Refs.\cite{Marian1,Marian2}).

\section{Elementary amplitude and $\Delta$ self-energy}

\subsection{Elementary amplitudes on- and off-shell}

The detailed information about the elementary reaction amplitude for
the free nucleon is given in Ref.\cite{UIM}. In the coherent $\pi^0$
photo- and electroproduction on spin zero nuclei only the spin
independent part contributes . Therefore, in the pion-nucleon $cm$
frame we have
\begin{equation}
{\tilde f}_{\gamma \pi}^{(\lambda)}=F_2({\tilde {\bf k}}_{\pi},
{\tilde {\bf k}}_{\gamma};W,Q^2) [\hat{\tilde {\bf k}}_{\pi}\times
\hat{\tilde {\bf k}}_{\gamma}] \cdot {\bf e}_{\lambda}\,,
\end{equation}
where $F_2$ is the standard CGLN amplitude and ${\tilde {\bf
    k}}_{\gamma (\pi)}$ is the photon (pion) momentum in the $\pi N$
$cm$ frame. These momenta can be boosted to an arbitrary frame by the
Lorentz transformation
\begin{equation}
{\tilde {\bf k}}_{\gamma (\pi)}={\bf k}_{\gamma (\pi)}+
\alpha_{\gamma (\pi)}{\bf P},\qquad
\alpha_{\gamma (\pi)}=\frac{1}{W_{i(f)}}\,\left(
\frac{{\bf P}\cdot {\bf k}_{\gamma (\pi)}}{E_{i(f)}+W_{i(f)}}
- \omega_{\gamma (\pi)} \right)\,,
\end{equation}
where $E_{i(f)}=\omega_{\gamma (\pi)}+E_N^{i(f)}$ and $W_{i(f)}$ are
the total energies for the photon-nucleon (pion-nucleon) systems in
the arbitrary and $cm$ frames, respectively. The total momentum ${\bf
  P}={\bf k}_{\gamma}+{\bf p}={\bf k}_{\pi}+{\bf p'}$ is given in the
arbitrary frame. Using the factorization approximation (13) and the
transformation (18), we can get a simple connection between the $f_2$
amplitude of Eq. (14) with the CGLN amplitude $F_2$,
\begin{equation}
f_2=\frac{k_{\gamma} k_{\pi}}{{\tilde k}_{\gamma} {\tilde k}_{\pi}}
\left[ 1+\frac{A-1}{2A}(\alpha_{\gamma}+\alpha_{\pi})\right]\,F_2\,.
\end{equation}

Within the UIM the amplitude $F_2$ contains contributions from Born and
vector meson exchange terms, $F_2^{(B)}$, and from "dressed" $N^*$
resonances, $F_2^{(R)}$.
In our present work we consider the photon's $lab$ energy range
$E_{\gamma}<400$ MeV, for which the s-channel $\Delta(1232)$ resonance
gives the most important contribution.  The multipole $M_{1+}^{(3/2)}$
related to the resonance is parametrized in a unitary way by a
standard Breit-Wigner form, i.e.
$F_2^{(R)}=2M_{1+}^{(\Delta)}+M_{1-}^{(Roper)}$, where
\begin{equation}
M_{1+}^{(\Delta)}(W,Q^2)\,=\,F_{\gamma N \Delta}(W,Q^2)
\frac{\Gamma_{\Delta}(W)\,M_{\Delta}\,e^{i\phi}}{M_{\Delta}^2-W^2-
iM_{\Delta}\Gamma_{\Delta}}\,F_{\pi N \Delta}(W)\,,
\end{equation}
and $M_{1-}^{(Roper)}$ accounts for the small contribution of the
Roper resonance $N^{\ast} (1440)$.  In Eq. (20), $F_{\gamma N \Delta}$
and $F_{\pi N \Delta}$ are the $\gamma N \Delta$ and $\pi N \Delta$
vertex function, respectively, $M_{\Delta}$ and $\Gamma_{\Delta}(W)$
the energy and total width of the $\Delta$ resonance as defined in
Ref.\cite{UIM}.  The unitary phase $\phi (W,Q^2)$ adjusts the phase of
the total multipole (background+resonance) to the corresponding
pion-nucleon scattering phase, in accordance with the Fermi-Watson
theorem.

As has been mentioned in the Introduction, it is one of the advantages
of the DWIA approach in momentum space that it allows us to account
for the nonlocality of the elementary operator and off-shell effects.
Such effects originate mainly from the principal value part of the Eq.
(16), the nonlocality being related with the dependence of the
amplitude on the pion momentum ${\bf k}_{\pi}$. For the background
contribution $F_2^{(B)}$, this dependence is in principle fixed by the
form of the Lagrangians.  However, the UIM background contribution is
derived from Lagrangians with a zero-range interaction.  In this case
the standard way (often used in coupled channels~\cite{Kaiser},
dynamical~\cite{NBL} or meson exchange~\cite{Yang} models) is to
introduce the finite range aspects of the interaction by an off-shell
form factor. In this spirit we have multiplied the background
contribution by a dipole-like form factor
\begin{eqnarray}
    f({\tilde k}_{\pi},{\tilde k}_{W})=\left(\frac{\Lambda^2 +
{\tilde k}_{W}^2}{\Lambda^2+{\tilde k}_{\pi}^{2}}\right)^2\,,
\end{eqnarray}
where ${\tilde k}_{W}$ is the on-shell value of the pion momentum in
the $\pi N$ $cm$ frame corresponding to the total energy $W$ for the
$\gamma N$ system. The standard value for the cut-off parameter
$\Lambda$, which provides the best fit in many calculations of
pion-nucleon scattering and pion photoproduction, is in the range of
400-500 MeV. In our present work we shall fix this parameter at
$\Lambda=450$ MeV.  For consistency with our off-shell extrapolation
of the pion-nucleon scattering amplitude in the $P_{33}$
channel~\cite{Marian1}, we shall use the prescription
\begin{eqnarray}
  M_{1+}^{(\Delta)}({\tilde k}_{\pi},W,Q^2)= M_{1+}^{(\Delta)}(W,Q^2)\,
  \frac{{\tilde k}_{\pi}}{{\tilde k}_{W}}\,
 f({\tilde k}_{\pi},{\tilde k}_{W})\,
\end{eqnarray}
for the $\Delta$ contribution, with the factor ${\tilde
  k}_{\pi}/{\tilde k}_{W}$ providing the correct threshold behavior in
the off-shell case.

\subsection{Self-energy of the $\Delta$}

It has been known from numerous $\Delta$-hole model calculations of
pion scattering, pion photoproduction and photoabsorption reactions
that the properties of the bound $\Delta$ isobar in the nuclear medium
differ substantially from those of the free $\Delta$ (see
Ref.~\cite{Weise} and references therein).

In the case of pion photo- and electroproduction, two main mechanisms
excite the $\Delta$ isobar excitation, i) a direct excitation with a
"bare" $\gamma N \Delta$ vertex, which corresponds to diagram (b) in
Fig. 1, and ii) a vertex renormalization mechanism, for which the
$\Delta$ is excited with pions produced by background terms (diagram
(c) in Fig.~1). As has been shown recently~\cite{KY} contributions of
these two diagrams are equally important and, therefore, medium
effects have to account for both mechanisms. The natural way to do so
is to introduce medium effects in the form of dressed resonance
contributions (diagram (a)) as defined by the UIM and given by Eq.
(20). However, as a first step, the relativistic $\Delta$ propagator
with the unitary phase $\phi$ has to be reduced to the nonrelativistic
Breit-Wigner form.  This can be achieved by the transformation
\begin{eqnarray}
\frac{e^{i\phi}}{M^2_{\Delta} - W^2 - i M_{\Delta}\Gamma_{\Delta}(W)}
= -\frac{1}{W+M_{\Delta}}\,\cdot\,
\frac{1}{W - {\bar M}_{\Delta}(W) + i{\bar \Gamma}_{\Delta}(W)/2}\,,
\end{eqnarray}
where
\begin{mathletters}
\begin{equation}
{\bar M}_{\Delta}(W) = W - (W-M_{\Delta})\cos\phi-
\frac{M_{\Delta}\Gamma_{\Delta}(W)}{W+M_{\Delta}}\sin\phi\,,
\end{equation}
\begin{equation}
{\bar \Gamma}_{\Delta}=\frac{2M_{\Delta}\Gamma_{\Delta}(W)}
{W+M_{\Delta}}\cos\phi  - 2(W-M_{\Delta})\sin\phi\,.
\end{equation}
\end{mathletters}

The next step is to modify DWIA amplitude by introducing the so-called
$\Delta$ self-energy $\Sigma_{\Delta}$ on the $rhs$ of Eq. (23), i.e.
\begin{eqnarray}
\frac{1}{W - {\bar M}_{\Delta} + i{\bar \Gamma}_{\Delta}(W)/2}
\rightarrow
\frac{1}{W - {\bar M}_{\Delta} + i{\bar \Gamma}_{\Delta}(W)/2
-\Sigma_{\Delta}}\,.
\end{eqnarray}

In the literature we find many approaches to calculate
$\Sigma_{\Delta}$ within simple models based on the local density
approximation~\cite{Oset1,Oset2} or more refined microscopical
calculations~\cite{Koch2,Koch1} based on the $\Delta$-hole model. Such
models were quite successful in describing pion scattering and pion
photoproduction on heavy nuclei. However, one of the peculiarities of
these approaches is that they finally always incorporate elements of
phenomenology. In our present work we will do so from the very
beginning by looking for a phenomenological parametrization of
$\Sigma_{\Delta}$ which should be simple, common for all nuclei and
able to describe the available data. For this purpose we shall test
two types of parametrization
\begin{mathletters}
\begin{equation}
\Sigma_{\Delta}(E_{\gamma},q^2)=V_1(E_{\gamma})\,F(q^2),\qquad
F(q^2)=e^{-\beta q^2}\,,
\end{equation}
and
\begin{equation}
\Sigma_{\Delta}(E_{\gamma},r)=(A-1)\,V_2(E_{\gamma})\,\rho(r)/\rho_0,\qquad
\rho_0=0.17 fm^{-3}\,,
\end{equation}
\end{mathletters}
where $V_{1,2}$ is a (complex and energy-dependent ) free parameter.
The first parametrization, Eq.~(26a), is convenient to use in momentum
space.  Here we assume that $F(q^2)$ is the $s$-shell harmonic
oscillator form factor with $\beta=0.54$ fm$^2$ and that
$\Sigma_{\Delta}$ is already saturated for $^4He$. This has the
consequence that $V_1$ and $F(q^2)$ have the same values for both
$^4He$ and $^{12}C$.  The second parametrization, Eq.~(26b), results
from the local density approximation, the nuclear density $\rho (r)$
being normalized to $\int \rho (r) d^3 r =1$.  Therefore, the $\Delta$
self-energy will differ for $^4He$ and $^{12}C$.

Finally we note that in the present context we neglect medium effects
due to the Roper resonance, because this contribution is very small in
our energy range, i.e. of the order of 2-3\% at a photon's $lab$
energy $E_{\gamma}< 400 $ MeV.

\section{Results and Discussion}

One of the attractive features of coherent pion photo- and
electroproduction on nuclei is that we can obtain clear signals from
modifications of the $\Delta$ isobar in the nuclear medium. For this
purpose, Fig. 2 presents our results for the reaction
$^4He(\gamma,\pi^0)^4He$ and $^{12}C(\gamma,\pi^0)^{12}C$ at a
photon's {\it lab} energy $E_{\gamma}=290$ MeV.  In this energy region
the standard DWIA calculations substantially overestimate the measured
differential cross sections. Our DWIA results (dashed curves) in the
maximum are larger than the experimental data by approximately a
factor of 2. As was shown in previous studies, one of the reasons for
this large discrepancy is the interaction of the $\Delta$ isobar with
the surrounding nucleons.

The $\Delta$-nucleus interaction leads to a renormalization of the
$\Delta$ propagator and can be described in terms of the $\Delta$
self-energy $\Sigma_{\Delta}$. The recently measured differential
cross sections for the reaction $^4He(\gamma,\pi^0)^4He$ and
forthcoming new data for $^{12}C(\gamma,\pi^0)^{12}C$ to be measured
over a wide energy region ($200 < E_{\gamma} < 400 $ MeV) provide us
with a unique opportunity to determine $\Sigma_{\Delta}$ with good
accuracy.  As has been pointed out in the last section, we shall now
test two types of parametrizations given by Eq. (26).

First, let us consider $\rho-type$ parametrization (26b)
conventionally used in the analysis of elastic pion-nucleus
scattering. Note that a Taylor expansion of Eq.~(25) in $\rho(r)$
leads to medium corrections in $\rho^2(r)$, which were often used in
analyses of pion-nucleus scattering.  One of the consequences of this
parametrization is that for finite nuclei it gives a good description
of the differential cross sections in forward direction but fail at
large angles and in the resonance region. On the other hand pion
scattering on light nuclei at backward angles is usually better
described without the phenomenological $\rho^2$ term (or without
$\Delta$ medium effects).

>From Fig. 2 we can see that the coherent $\pi^0$ photoproduction
situation is similar to the case described above (see the dash-dotted
curves). This result clearly indicates that for light nuclei the
treatment of the $\Delta$-nucleus interaction in terms of $\rho(r)$
becomes less satisfactory.  Therefore, nuclei like $^4He$ require a
more sophisticated microscopical calculation of the $\Delta$
self-energy, e.g. the $\Delta$-hole~\cite{Koch2,Koch1} or meson
exchange models~\cite{Oset3}. We would like to stress that the first
microscopical $\Delta$-hole calculation performed some 15 years ago by
J.  Koch and E. Moniz~\cite{Koch1} gives excellent agreement with the
results of the recent measurements (see the dotted curve in Fig. 2).

However, a consistent extension of the microscopical calculations to
heavier nuclei meets serious difficulties, and at some level it always
requires elements of phenomenology. Therefore, we decide to use
another approach in our present work. First, with the use of the
$F-type$ parametrization (26a) in momentum space, we extracted
$\Sigma_{\Delta}$ from the data for the reaction
$^4He(\gamma,\pi^0)^4He$. Then, we assume that the $\Delta$-nucleus
interaction is the same also for the heavier nuclei, i.e., that it
saturates already for $^4He$. Our calculations at $E_{\gamma}=290$ MeV
indicate that this is indeed a realistic assumption, and allows us to
describe the coherent $\pi^0$ photoproduction on $^4He$ and $^{12}C$
over a wide range of pion angles (see the solid curves in Fig. 2). The
obtained values for the strength of the $\Delta$-nucleus interaction,
$Re\ V_1$=19 MeV and $Im\ V_1$= -33 MeV, are in reasonable agreement
with the values obtained in pion-nucleus scattering.

This result inspires us to test our model at other energies
assuming that $V_1$ is an energy dependent function to be
determined by fitting the experimental data for $^4He$. In Fig. 3
we present results of our fit in the energy range $200 MeV <
E_{\gamma} < 400$ MeV. In Fig. 4 the prediction of our model
is compared with the new experimental data from Ref.\cite{BNL}.
The energy dependence for the total cross section and the
parameters $V_1$ are shown in Fig. 5. These results for the
potential parameter $V_1$ agree qualitatively well with the
predictions of Ref.\cite{Sauer}.

Our most interesting result is shown in Fig. 6, where the prediction
is compared to the data of the A2 collaboration at MAMI for the
reaction $^{12}C(\gamma,\pi^0)^{12}C$ obtained at a pion angle
$\theta_{\pi}=60^0$. Even at such a large angle where the differential
cross section has dropped by more than an order of magnitude (see Fig.
2), our assumption about the saturation of the $\Delta$-nucleus
interaction is still quite acceptable. Of course, a better
understanding of this phenomenon will require additional data at
smaller angles where the differential cross section reaches a maximum,
as well as data for other nuclei.

A new feature of the $\Delta$-nucleus interaction can be investigated
and new information about the $Q^2$ dependence of $\Sigma_{\Delta}$
can be obtained by using virtual photons (or coherent $\pi^0$
electroproduction). Clearly, as the $\Delta$ isobar should not
remember how it was excited, the $V_1$ does not depend on $Q^2$.
Therefore, the strength of its interaction with the surrounding
nucleons is universal for all processes (electroproduction, pion
scattering, Compton scattering, etc.) and the $Q^2$ dependence enters
the parametrization of Eq.~(26a) only in the form factor $F(q)$, with
$ q=|{\bf k}_{\gamma}-{\bf k}_{\pi}|$ the momentum transferred to the
nucleus and ${\bf k}_{\gamma}$ the virtual photon momentum as function
of $Q^2$,
\begin{equation}
{\bf k}^2_{\gamma}=Q^2+\frac{(s_A-M_A^2-Q^2)^2}{4s_A}=
Q^2+\omega_{\gamma}^2\,.
\end{equation}

In Fig. 7 we depict the predictions of our model for: i) an equivalent
photon energy $E_{\gamma}$=228 MeV and $Q^2$=0, 0.054, 1.81
$(GeV/c)^2$, and ii) $E_{\gamma}$=289 MeV and $Q^2$=0, 0.074, 1.61
$(GeV/c)^2$. This corresponds to the kinematics of the recent and
forthcoming data from NIKHEF~\cite{Block}

\section{Conclusion}

In this paper we have presented a first nuclear application of our
recently developed unitary isobar model to pion photo- and
electroproduction.  This model describes all the existing data for
$(\gamma,\pi)$ and $(e,e'\pi)$ on the nucleon reasonably well and
furthermore compares very well with the partial wave analysis of the
VPI group up to $W_{cm}=1700 MeV$.

We have performed a DWIA calculation for coherent $\pi^0$ photo- and
electroproduction from nuclei and applied it to experimentally
investigated reactions on $^4He$ and $^{12}C$. Due to the very precise
data recently measured at Mainz over a large angular range of photon
$lab$ energies from 207 MeV up to 397 MeV, we are able to determine,
for the first time, the energy dependence of the $\Delta$-nucleus
potential from pion photoproduction reactions. Comparing $^4He$ and
$^{12}C$ we find no A-dependence of the potential and conclude that
the $\Delta$-nucleus interaction saturates already for $^4He$ .

In our study of the dynamical mechanism of the $\Delta$ self-energy in
the nuclear medium we find that a form factor type $\Delta$-nucleus
interaction is preferred as compared to a density type of medium
modification often used in the local density approximation of
coordinate space calculations.

First experimental results on electroproduction obtained at NIKHEF are
in good agreement with our calculations and no readjustment of the
$\Delta$-nucleus interaction is needed.

For future experiments it will be very interesting to see how the
$\Delta$-nucleus interaction saturates by measurements on the deuteron
and on $^3He$ and, secondly, if the saturation observed for $^{12}C$ is
valid for the heavier nuclei.

\acknowledgements

We would like to thank M. Kirchbach, F. Wissman, F. Rambo, H. Blok,
T. Botto and D. H. Lu for helpful discussions.  S.S.K. and L.T. are
grateful to the Physics Department of the NTU for the hospitality
extended to them during their visit. This work was supported in part
by the Deutsche Forschungsgemeinschaft (SFB 443) and by the National
Science Council/ROC under grant NSC 88-2112-M002-015.

\begin{figure}[h]
\centerline{\epsfig{file=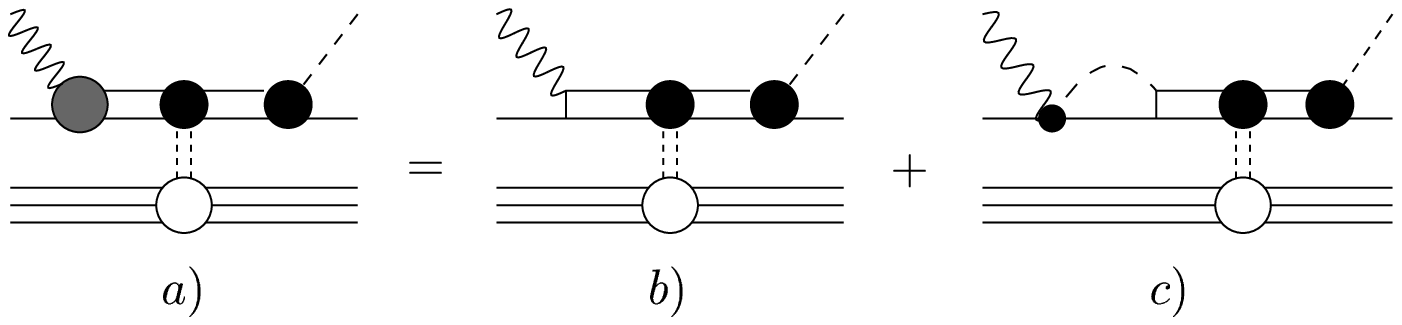, width=12 cm} }
\caption{ Two main mechanisms of the $\Delta$ isobar excitation and
corresponding medium effects: b) direct excitation with "bare"
$\gamma N \Delta$ vertex; c) vertex renormalization mechanism
where $\Delta$ isobar is excited from pions produced by
nonresonant background.}
\end{figure}

\begin{figure}[h]
\epsfig{file=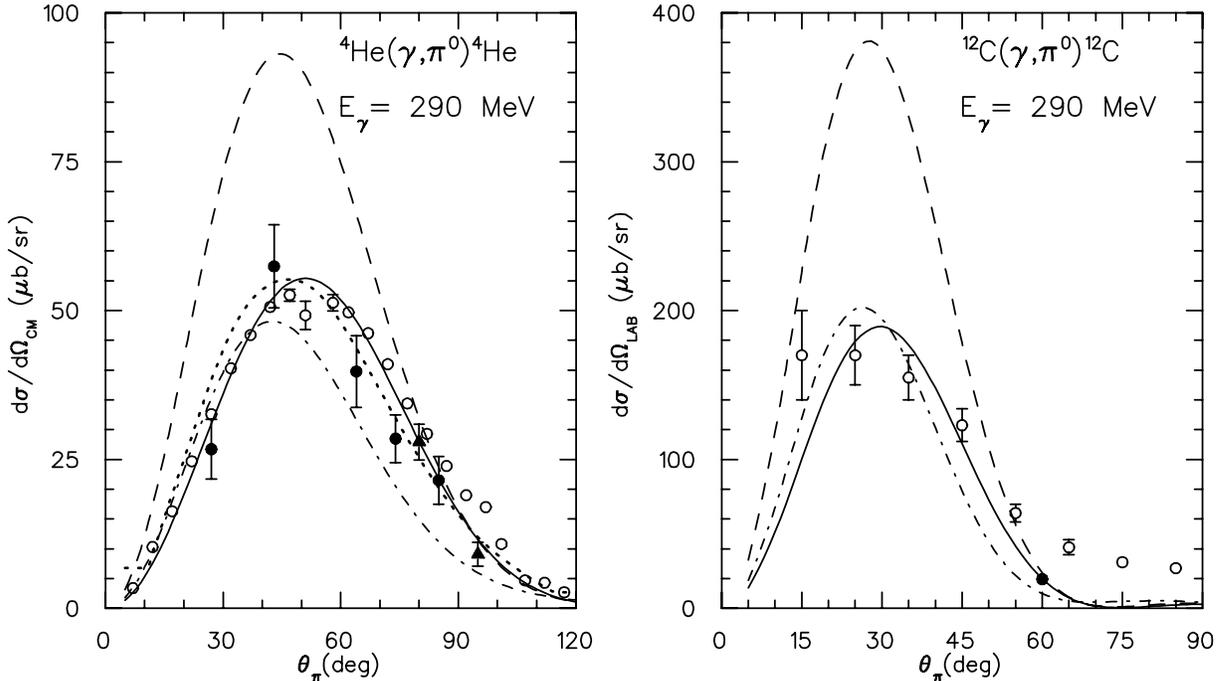, width=9 cm, angle= 90}
\caption{
  The differential cross sections for the $^4He(\gamma,\pi^0)^4He$ and
  $^{12}C(\gamma,\pi^0)^{12}C$ reactions at photon $lab$ energy
  $E_{\gamma}=290$ MeV. Dashed curves are the DWIA results.
  Dash-dotted and solid curves are the results obtained with
  $\rho-type$ (26b) and $F-type$ (26a) parametrizations for the
  $\Delta$ self-energy. Dotted curve is the $\Delta$-hole model
  calculation taken from Ref.\protect\cite{Koch1} and recalculated in
  the $\pi ^4$He c.m. frame. Experimental data for $^4$He are from
  Refs.\protect\cite{Rambo} (open circles), \protect\cite{Exp1} (full
  circles) and \protect\cite{Exp2} (full triangles). Experimental
  data for $^{12}$C are from Refs.\protect\cite{Arends} (open circles)
  and \protect\cite{A2} (full circles).  }
\end{figure}

\begin{figure}[h]
\epsfig{file=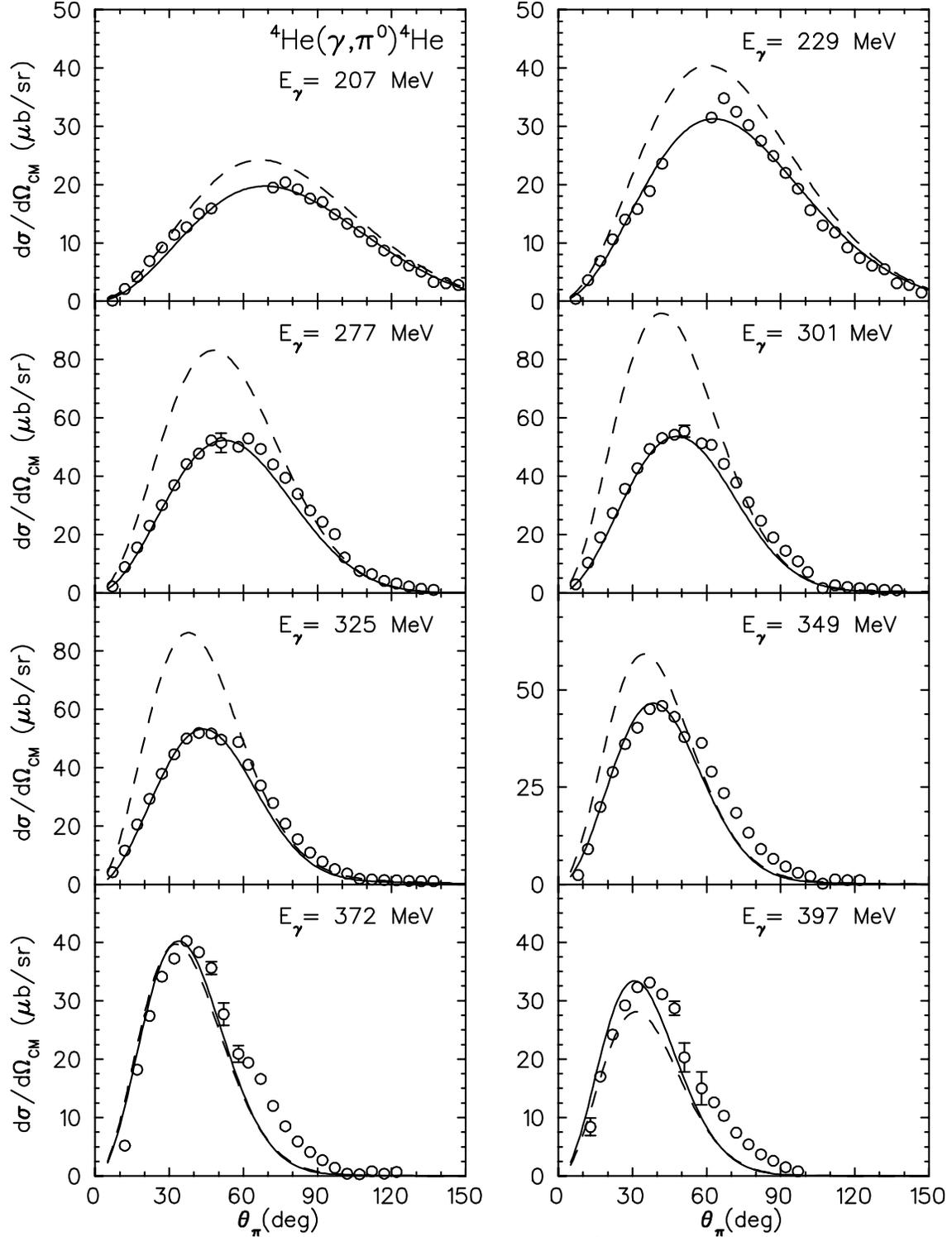, width=15 cm} \caption{ The differential
cross sections for the $^4He(\gamma,\pi^0)^4He$ reaction. Dashed
curves are the DWIA results. Solid curves are the results obtained
with $F-type$ (26a) parametrizations for the $\Delta$ self-energy.
Experimental data are from Ref.\protect\cite{Rambo}. }
\end{figure}

\begin{figure}[h]
\epsfig{file=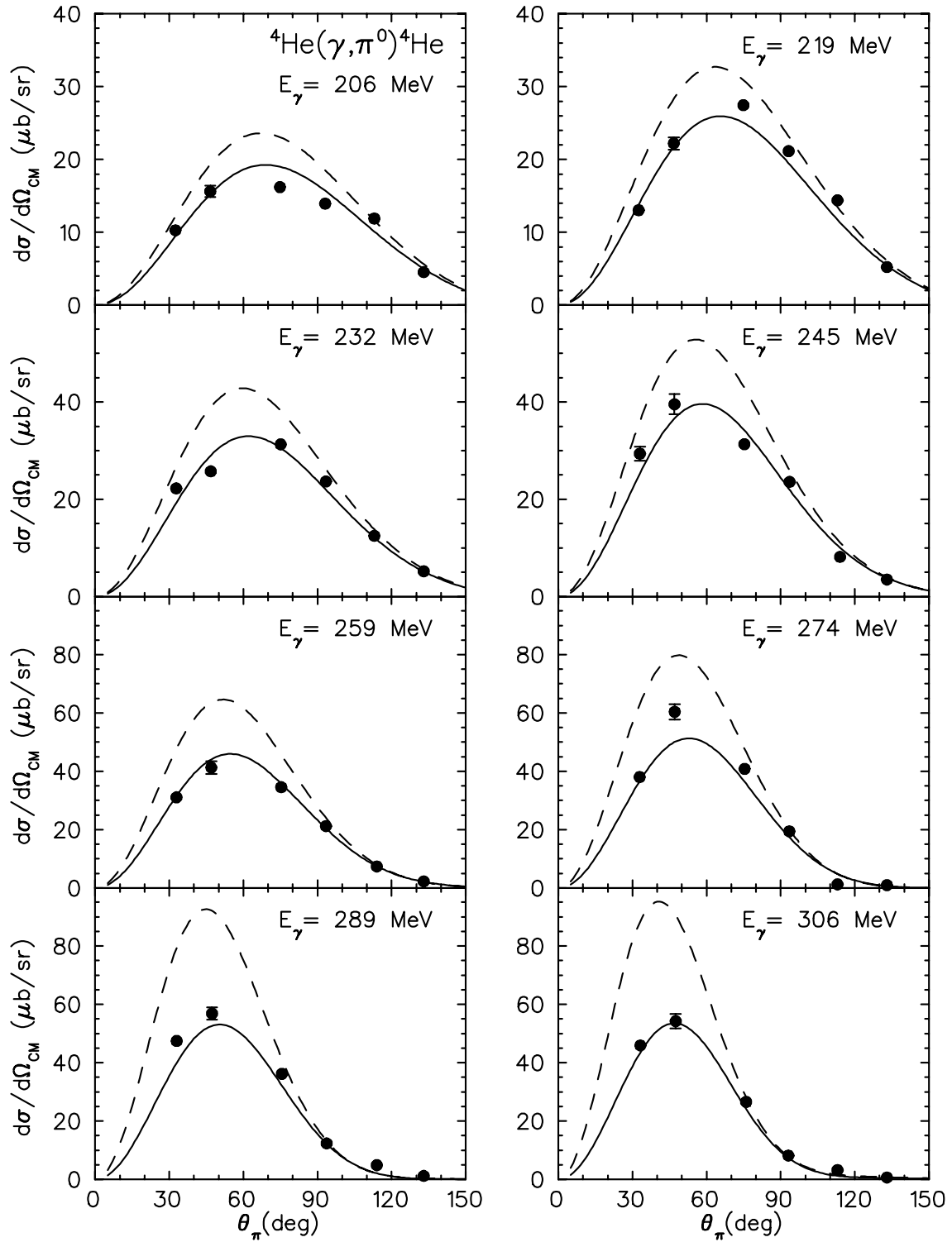, width=15 cm} \caption{ The same as in Fig.
3. Experimental data are from Ref.\protect\cite{BNL}. }
\end{figure}

\begin{figure}[h]
\epsfig{file=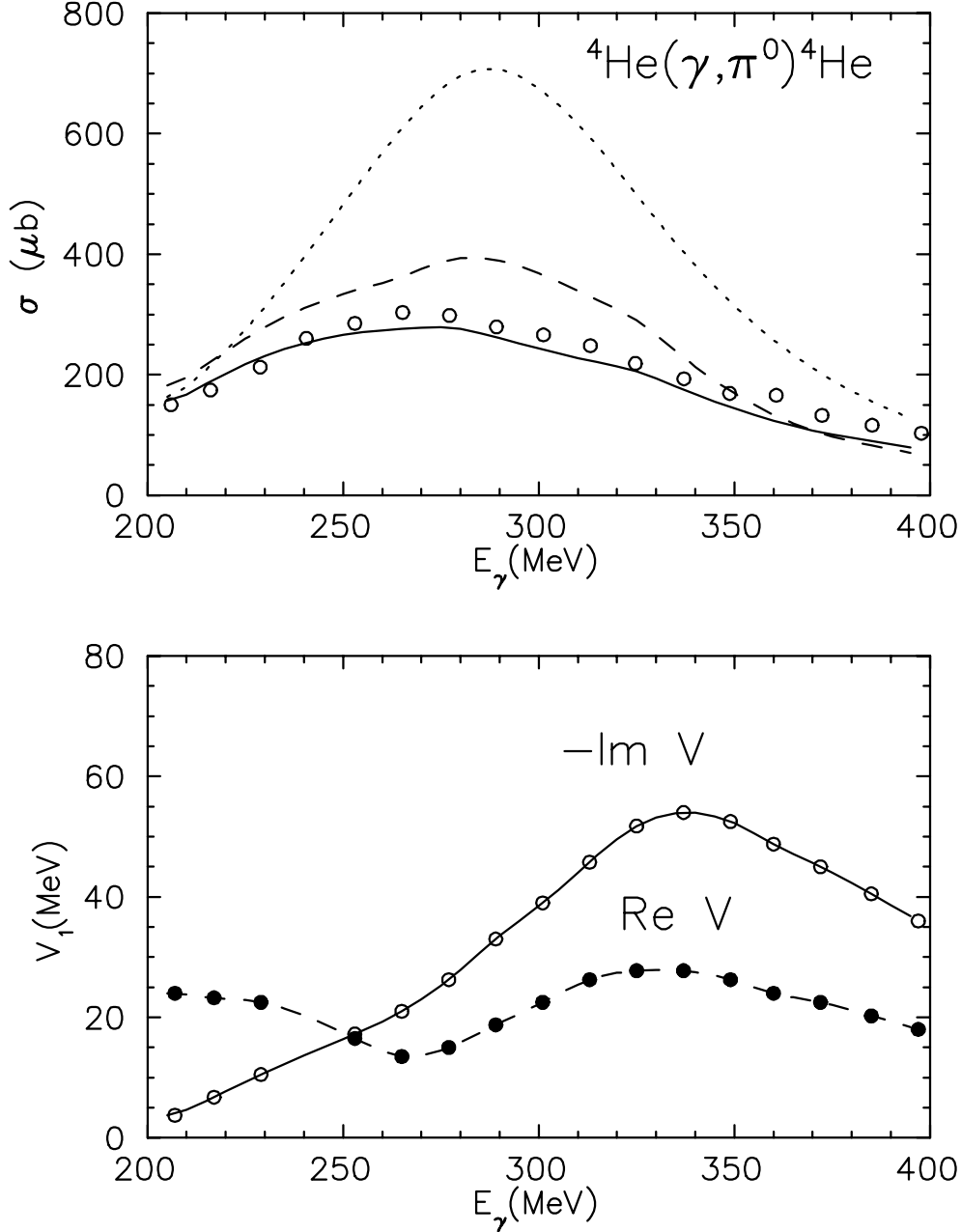, width=13 cm}
\caption{ Energy dependence of the total cross section for the
$^4He(\gamma,\pi^0)^4He$ reaction (upper figure)
and the corresponding values of the $F-type$ $\Delta$-self energy
$V_1(E_{\gamma})$  from Eq. (26a) (lower figure). In the upper figure,
dotted and dashed curves are the total cross sections for PWIA and DWIA,
respectively. The solid curve is the result obtained with $F-type$ $\Delta$
self-energy. Experimental data are from Ref.\protect\cite{Rambo}.
}
\end{figure}

\begin{figure}[h]
\epsfig{file=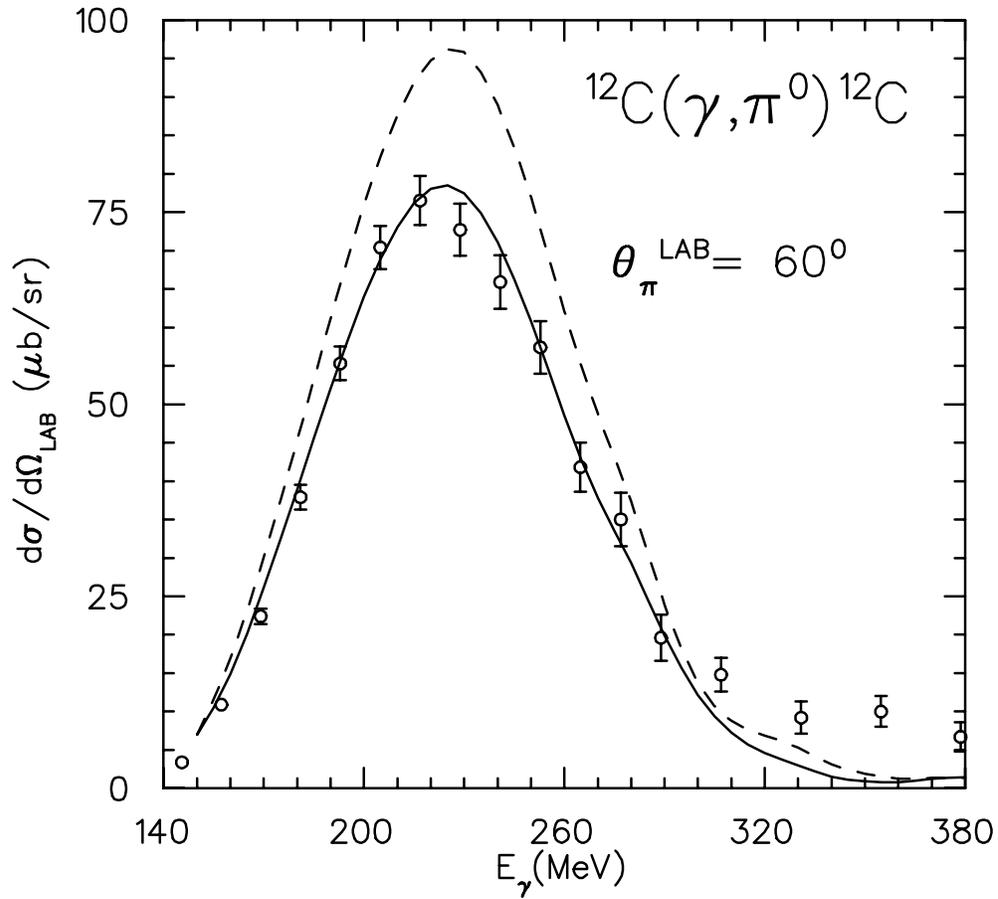, width=13 cm }
\caption{ Energy dependence of the differential cross section for the
$^{12}C(\gamma,\pi^0)^{12}C$ reaction at pion $lab$ angle
$\theta_{\pi}=60^0$. Notation for the curves  as in Fig. 3.
Experimental data are from Ref.\protect\cite{A2}.
}
\end{figure}

\begin{figure}[h]
\epsfig{file=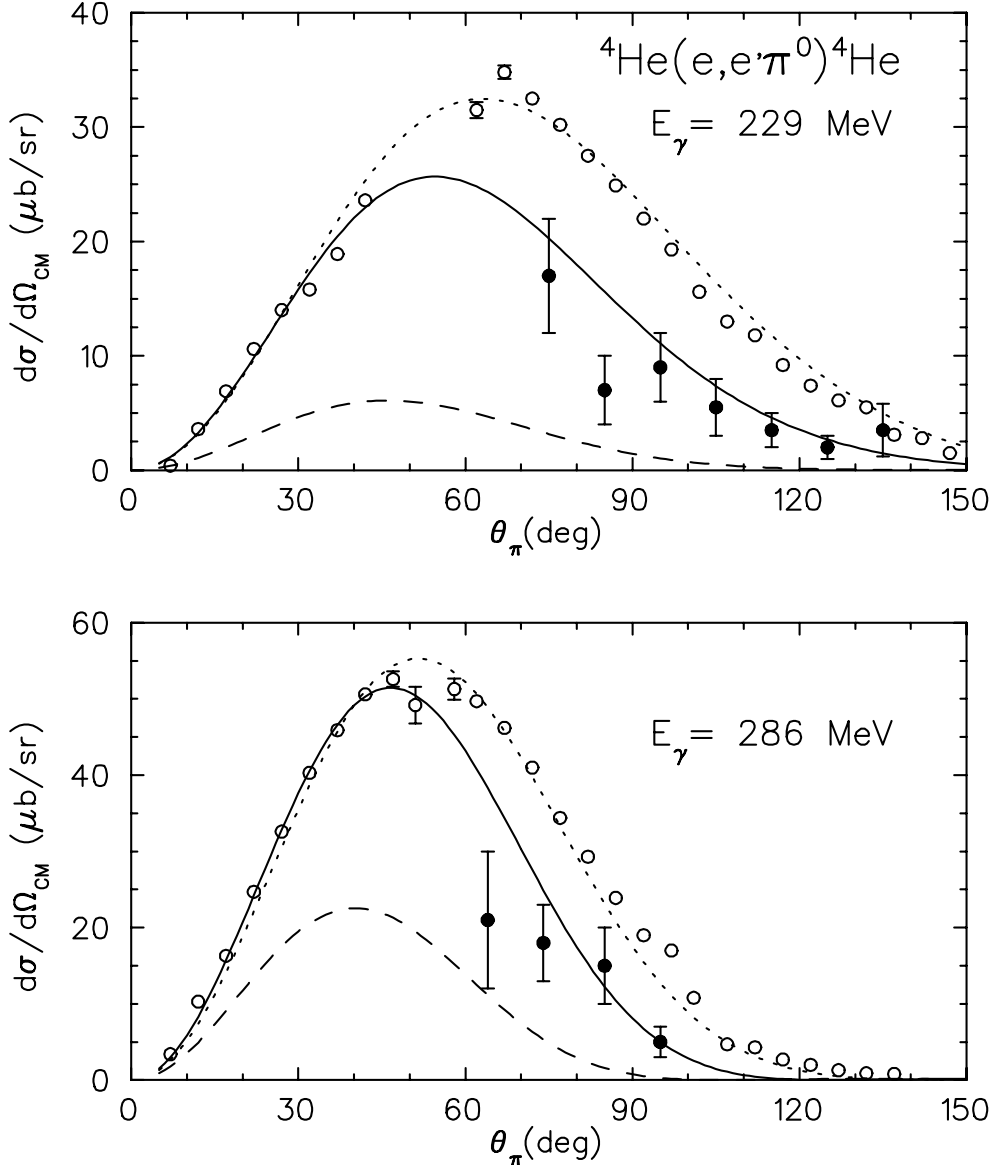, width=13 cm}
\caption{ Differential cross section for the coherent $\pi^0$
electroproduction on $^4$He calculated with $F-type$ $\Delta$ self energy,
Eq. (26a). The dotted curves are the results at the photon point ($Q^2=0$).
The corresponding experimental data are from Ref.\protect\cite{Rambo}
(open circles).
The solid, and dashed curves in the upper (lower) figure are the results for
$Q^2$=0.062 (0.054) and 1.81 (1.60) $(GeV/c)^2$, respectively.
The experimental data~\protect\cite{Block} (full circles) correspond
to $Q^2=0.062 (GeV/c)^2$ in the upper figure and $Q^2=0.054 (GeV/c)^2$
in the lower figure.
}
\end{figure}

\end{document}